\documentclass[12pt]{iopart}

\usepackage{graphicx}
\usepackage{cite}
\usepackage{iopams}

\usepackage{amstext}

\begin{document}

\title[Short coherent control pulse with small random errors in its direction]{Short coherent control pulse with small random errors in its direction}

\author{Su Zhi-Kun and Jiang Shao-Ji}

\address{State Key Laboratory of Optoelectronic Materials and Technologies, Sun Yat-sen University, Guangzhou 510275, People's Republic of China}
\ead{stsjsj@mail.sysu.edu.cn}

\begin{abstract}
The finite-amplitude short coherent control pulse with small random errors in its direction is considered. We derive the conditions for
it to approximate an ideal $\delta$-shaped pulse up to an error in the second order of pulse duration. Whether the symmetric and asymmetric pulse shapes would fulfill these
conditions is also analyzed. The result shows that, under one of these conditions, all symmetric pulse shapes can fulfill this condition but not for all asymmetric ones.

\end{abstract}

\maketitle

\section{Introduction}

Suppressing decoherence is significantly important for quantum information
processing (QIP) and other quantum-based technologies. One of the most
effective techniques to combat with the decoherence process is dynamical
decoupling (DD)\cite{PhysRevA.58.2733,PhysRevLett.82.2417}, which evolves from the
Hahn echo\cite{PhysRev.80.580} and develops for\ refocusing techniques in
nuclear magnetic resonance (NMR)\cite{PhysRevLett.25.218,Mehring1983,Haeberlen1976}. The
basic idea of DD\cite{PhysRevB.77.174509,jpb.44.154002,jpb.45.025502} is to insert a
sequence of short coherent control pulses within a time period of interest,
such that the system of interest can be effectively decoupled from its
environment. In the theoretical treatments the control pulse is considered as
an ideal, i.e. $\delta$-shaped $\pi$ pulse (so-called ``bang-bang" or unbounded
control) which means that its amplitude is infinite in the sense of a $\delta$
function. This assumption is convenient because all the Hamiltonian of the
system can be neglected during the control pulse duration (infinitely short).
Hence it is relatively straightforward to cope with the time evolution.

In experiments, however, the ideal instantaneous pulse is not
realistic\cite{Nature.458.996,PhysRevA.82.042306} since the pulse would have a finite
duration $\tau_{p}$ or it might be imperfect, e.g., one can have pulse length
or amplitude errors (leading to a wrong angle of rotation) or an off-resonance
error, due to which the rotation occurs around a tilted axis. The accumulation
effect of these small imperfections in the pulse may have significant consequences
when the number of pulses in a DD pulse sequence is not small. The issue of finite duration and finite amplitude has been considered in \cite{PhysRevA.77.032315,jpa.41.312005,PhysRevA.78.022315,PhysRevA.83.022306}. To mitigate
the effect of finite duration and finite amplitude, pulse shaping has been addressed to approximate an instantaneous pulse about a fixed duration\cite{PhysRevA.77.032315}. Following the idea developed in \cite{PhysRevA.77.032315}, the
case of a fixed axis has been extended to the case of an axis varying in
time\cite{jpa.41.312005}. Further numerical work illustrates that tailoring the
pulse, including symmetric and asymmetric, improves the quality of a real
pulse significantly\cite{PhysRevA.78.022315}. But the small deviations of the actual direction of the pulse have not been considered in \cite{PhysRevA.77.032315,jpa.41.312005,PhysRevA.78.022315,PhysRevA.83.022306}.

On the other hand, the effect of small random errors in the direction has been studied in a comprehensive comparison of various DD sequence in \cite{PhysRevA.83.032322,jpb.44.154004}. However, the $\pi$ pulse considered in \cite{PhysRevA.83.032322,jpb.44.154004} is instantaneous and its amplitude is infinite. Hence the discussion is given in the absence of the coupling to the bath during the pulse and on the level of pulse sequences. In this paper, we consider a finite-amplitude pulse with random errors in its directions on the level of an individual pulse. We aim to provide some guidelines in the choice of pulse
shapes, i.e. symmetric and asymmetric shapes, when the rotation occurs around a
tilted axis. To realize the goal, we follow the idea proposed in
\cite{PhysRevA.83.022306} and derive the expressions of error terms for this finite-amplitude pulse to
simulate an ideal pulse up to an error of $O(\tau_{p}^{2})$ in the presence of small random errors in its direction, where $\tau_{p}$ is the pulse duration. We divide these terms into two groups, those from pulse duration only and the
others originate from errors in pulse direction, and discuss them separately.
As expected, both symmetric and asymmetric pulse shapes can make the error
terms from pulse duration vanish. However, neither type of the pulse can
remove the error terms from pulse direction completely, but all symmetric
pulse can make one more terms equal zero than the specific choice of asymmetric one. Our
results can be useful for experimental implementations of kinds of DD
protocols, and for deeper understanding of a short coherent
control pulse which is subject to random errors around its direction.

The paper is organized as follows. After introducing the model (section 2), we
start discussing the error terms due to pulse duration and reviewing two kinds
of optimized pulse shapes which are designed to eliminate these terms (section
3). Then the error terms coming from deviation in rotation axis are analyzed
(section 4). At last the conclusion is
given (section 5).

\section{Model}

The total Hamiltonian of system, bath and the interaction between
them is given by $H$. To suppress decoherence, we implement fast control on
the system and the control Hamiltonian reads

\begin{equation}
H_{\Omega}(t)=V(t)\Omega,\label{HO}%
\end{equation}
where $V(t)$ stands for the pulse shape as a function of time and $\Omega$ is
a parity kick operator, i.e., $\Omega\Omega^{\dag}=\Omega^{2}=1$\cite{PhysRevA.59.4178}.
In reality, the controlling pulses may be subject to small errors in its
direction, so that the the Hamiltonian $H_{\Omega}(t)$ in Eq.(\ref{HO}%
)\ should be corrected into

\begin{equation}
H_{C}(t)=V(t)\left(  \Omega+\epsilon\Omega^{\prime}\right)  ,\label{HC}%
\end{equation}
where $\epsilon$ is the small error factor and $\Omega^{\prime}$ is
an arbitrary operator of the system. The evolution operator during the pulse
of length $\tau_{p}$ takes the form of

\begin{eqnarray}
U(\tau_{p},0)  & =\overleftarrow{T}\exp\left\{  -i\int\nolimits_{0}^{\tau_{p}%
}\left[  H+H_{C}(t)\right]  dt\right\} \nonumber\\
& =e^{-iH(\tau_{p}-\tau_{s})}U_{C}e^{-iH\tau_{s}},\label{U}%
\end{eqnarray}
where $\overleftarrow{T}$ stands for the time-ordering operator with time
increasing from right to left, $\tau_{s}$ is an arbitrary instant inside the
interval $\left[  0,\tau_{p}\right]  $ (more detail about will be given below
Eq.(\ref{UN})), and
\begin{equation}
U_{C}=\overleftarrow{T}\exp\left[  -i\int\nolimits_{0}^{\tau_{p}}\left[
\widetilde{H}_{C}(t)\right]  dt\right] \label{UC}%
\end{equation}
with $\widetilde{H}_{C}(t)=e^{iH(t-\tau_{s})}H_{C}(t)e^{-iH(t-\tau_{s})}$. On
the other hand, the evolution, described by Eq.(\ref{U}), can be rewritten as

\begin{eqnarray}
U(\tau_{p},0)  & =\overleftarrow{T}\exp\left\{  -i\int\nolimits_{0}^{\tau_{p}%
}\left[  H+H_{C}(t)\right]  dt\right\} \nonumber\\
& =e^{-iH(\tau_{p}-\tau_{s})}P_{\Omega}e^{-iH\tau_{s}}+O(\epsilon
)+O(\epsilon\tau_{p})+O(\tau_{p}^{2}),\label{UN}%
\end{eqnarray}
where $P_{\Omega}=\exp\left[  -i\int\nolimits_{0}^{\tau_{p}}H_{\Omega
}(t)dt\right]  $ is the desired instantaneous control applied at the time
$\tau_{s}$. Next, we follow the method outlined in \cite{PhysRevA.83.022306} to find out
the expressions of the error terms $O(\epsilon)+O(\epsilon\tau_{p})$ in the
last line of Eq.(\ref{UN}). Compare $U_{C}$ in Eq.(\ref{U}) with $P_{\Omega}$
in Eq.(\ref{UN}), we evaluate first the error in the control Hamiltonian
\begin{eqnarray}
h_{\Omega}(t)  & =\widetilde{H}_{C}(t)-H_{\Omega}\nonumber\\
& =V(t)\left\{  \sum\limits_{k=1}^{\infty}\frac{(t-\tau_{s})^{k}}{k!}\left[
\left(  iH\right)  ^{k},\Omega\right]  +\epsilon\Omega^{\prime}+\epsilon
\sum\limits_{k=1}^{\infty}\frac{(t-\tau_{s})^{k}}{k!}\left[  \left(
iH\right)  ^{(k)},\Omega'\right]  \right\}  .\label{homiga}%
\end{eqnarray}

The notation stands for%

\begin{equation}
\left[  (iH)^{(1)},\cdots\right]  =\left[  iH,\cdots\right]  ,\label{na}%
\end{equation}

\begin{equation}
\left[  (iH)^{(2)},\cdots\right]  =\left[  iH,\left[  iH,\cdots\right]
\right]  ,\label{nb}%
\end{equation}

\begin{equation}
\left[  (iH)^{(3)},\cdots\right]  =\left[  iH,\left[  iH,\left[
iH,\cdots\right]  \right]  \right]  ,\label{nc}%
\end{equation}
and so on. Now we have the error in the evolution
\begin{eqnarray}
\delta P_{\Omega}  & \equiv U_{C}-P_{\Omega}\nonumber\\
& =\overleftarrow{T}\{e^{-i\int\nolimits_{0}^{\tau_{p}}H_{\Omega}%
(t)dt}[e^{-i\int\nolimits_{0}^{\tau_{p}}h_{\Omega}(t)dt}-1]\}.\label{deltaP}%
\end{eqnarray}

Then the leading-order in $\delta P_{\Omega}$, i.e. to the first
order in $\tau_{p}H$ and $\epsilon$, is\cite{PhysRev.84.108}

\begin{eqnarray}
\eta & =\overleftarrow{T}\left\{  e^{-i\int\nolimits_{0}^{\tau_{p}}H_{\Omega
}(t)dt}\int\nolimits_{0}^{\tau_{p}}V(t)\{(t-\tau_{s})[H,\Omega]-i\epsilon
\Omega^{\prime}+(t-\tau_{s})\epsilon\lbrack H,\Omega^{\prime}]\}dt\right\}
\nonumber\\
& =\int\nolimits_{0}^{\tau_{p}}V(t)e^{-i\int\nolimits_{t}^{\tau_{p}}H_{\Omega
}(s)ds}\{-i\epsilon\Omega^{\prime}+(t-\tau_{s})\epsilon\lbrack H,\Omega
^{\prime}]+(t-\tau_{s})[H,\Omega]\}e^{-i\int\nolimits_{0}^{t}H_{\Omega}%
(s)ds}dt\nonumber\\
& =\eta^{(\epsilon,1)(\tau_{p},0)}+\eta^{(\epsilon,1)(\tau_{p},1)}+\eta
^{(\tau_{p},1)},\label{epsilon}%
\end{eqnarray}
where the superscript $\left(  u,v\right)  $ of $\eta$\ in the last line of
Eq.(\ref{epsilon})\ denotes an error in the $v$th order of $u$.\ Take the
superscript $\eta^{(\epsilon,1)(\tau_{p},1)}$\ for an example, this term means
an error in the first order of $\epsilon$ and an error in the first order of
$\tau_{p}$\ as well. Here we have defined
\begin{equation}
\eta^{(\epsilon,1)(\tau_{p},0)}=-i\epsilon\int\nolimits_{0}^{\tau_{p}%
}V(t)e^{-i\int\nolimits_{t}^{\tau_{p}}H_{\Omega}(s)ds}\Omega^{\prime}%
e^{-i\int\nolimits_{0}^{t}H_{\Omega}(s)ds}dt,\label{e1}%
\end{equation}
\begin{equation}
\eta^{(\epsilon,1)(\tau_{p},1)}=\epsilon\int\nolimits_{0}^{\tau_{p}}%
(t-\tau_{s})V(t)e^{-i\int\nolimits_{t}^{\tau_{p}}H_{\Omega}(s)ds}%
[H,\Omega^{\prime}]e^{-i\int\nolimits_{0}^{t}H_{\Omega}(s)ds}dt,\label{e2}%
\end{equation}
and%
\begin{equation}
\eta^{(\tau_{p},1)}=\int\nolimits_{0}^{\tau_{p}}(t-\tau_{s})V(t)e^{-i\int
\nolimits_{t}^{\tau_{p}}H_{\Omega}(s)ds}[H,\Omega]e^{-i\int\nolimits_{0}%
^{t}H_{\Omega}(s)ds}dt.\label{e3}%
\end{equation}

Note that the term described by Eq.(\ref{e3}) is the same as those derived in
\cite{PhysRevA.83.022306,PhysRevA.77.032315}. The error $\eta^{(\tau_{p},1)}$ does not come
from directional error and it can be eliminated using pulse shaping
\cite{PhysRevA.77.032315,jpa.41.312005,PhysRevA.78.022315}. In the following two sections, we
will analyze how shaped pulses reduce the errors$\eta^{(\tau_{p},1)}$,
$\eta^{(\epsilon,1)(\tau_{p},0)}$, and $\eta^{(\epsilon,1)(\tau_{p},1)} $,
respectively, and with which types of pulse shape the errors will approximate
as close as possible to zero. Furthermore, we should mention that, the model
discussed in this section is general, for example, we can take $\Omega
\rightarrow\sigma_{x}$ and $\epsilon\Omega^{\prime}\rightarrow\epsilon
_{y}\sigma_{y}+\epsilon_{z}\sigma_{z}$ for a single-qubit system and the
control pulse corresponds to a rotation about $x$ axis with small deviation
$\epsilon_{y}$ and $\epsilon_{z}$ in the $y$ and $z$ direction, respectively,
where $\sigma_{x}$, $\sigma_{y}$ and $\sigma_{z}$ are the standard Pauli matrices.

\section{Error due to pulse duration ($\eta^{(\tau_{p},1)}$)}

In this section, the expression of $\eta^{(\tau_{p},1)}$\ is given, then the
symmetry and asymmetric pulse shapes, which are designed to eliminate
$\eta^{(\tau_{p},1)}$, are also introduced.

We begin to calculate the expression. Using $\Omega^{2}=1$ and Eq.(\ref{HO}), we have%

\begin{equation}
e^{-i\int\nolimits_{t_{1}}^{t_{2}}H_{\Omega}(s)ds}=\cos[\int\nolimits_{t_{1}%
}^{t_{2}}V(s)ds]-i\Omega\sin[\int\nolimits_{t_{1}}^{t_{2}}%
V(s)ds].\label{formula}%
\end{equation}

The total Hamiltonian $H$ can be decomposed into two parts\cite{PhysRevA.83.022306}%

\begin{equation}
H=H_{a}+H_{c},
\end{equation}
with $H_{a}$ and $H_{c}$\ anticommuting and commuting with
$\Omega$, respectively. Then from Eq.(\ref{e3}) one has%

\begin{equation}
\eta^{(\tau_{p},1)}=i2H_{a}\eta_{1}^{(\tau_{p},1)}+2H_{a}\Omega\eta_{2}%
^{(\tau_{p},1)},\label{e3e}%
\end{equation}
where%

\begin{equation}
\eta_{1}^{(\tau_{p},1)}=\int\nolimits_{0}^{\tau_{p}}(t-\tau_{s})V(t)\sin
\left[  \phi_{-}-\psi(t)\right]  dt,\label{e3ea}%
\end{equation}

\begin{equation}
\eta_{2}^{(\tau_{p},1)}=\int\nolimits_{0}^{\tau_{p}}(t-\tau_{s})V(t)\cos
\left[  \phi_{-}-\psi(t)\right]  dt,\label{e3eb}%
\end{equation}
with $\psi(t)\equiv2\int\nolimits_{\tau_{s}}^{t}V(s)ds$ and $\phi_{\pm}%
=\int\nolimits_{\tau_{s}}^{\tau_{p}}V(s)ds\pm\int\nolimits_{0}^{\tau_{s}%
}V(s)ds$ ($\phi_{+}$ will be used later). To make $\eta_{1}^{(\tau_{p}%
,1)}=\eta_{2}^{(\tau_{p},1)}=0$, two kinds of composite pulses, which consist
of piecewise constant pulses of maximally positive or negative amplitude $\pm
a_{\max}$, have been designed in \cite{PhysRevA.77.032315,jpa.41.312005,PhysRevA.78.022315}.
Here we review so-called symmetric pulses first and then asymmetric pulses.

For symmetric pulses with $\tau_{s}=\tau_{p}/2$, one needs at least two free
parameters including $\tau_{s}$, in order to obtain pulses which show only
quadratic deviations from the idealized instantaneous pulse. One way to
approach the solution of the one additional free parameter is to take the
instant $\tau_{1}$, at which the pulse changes first and then another sign
change occurs by symmetry at $t=\tau_{p}-\tau_{1}$. The pulse shape $V(t)$ can
be depicted as\bigskip%
\begin{equation}
V_{s}(t)=\left\{
\begin{array}
[c]{ccccc}%
a_{\max} & if & \tau_{1}\leq t\leq\tau_{p}-\tau_{1} &  & \\
-a_{\max} & if & 0\leq t<\tau_{1} & or & \tau_{p}-\tau_{1}<t\leq\tau_{p}%
\end{array}
\right.  .\label{vs}%
\end{equation}

\bigskip For asymmetric pulses case, on the other hand, the composite pulse
consists of two constant regions only. The pulse shape gives%

\begin{equation}
V_{a}(t)=\left\{
\begin{array}
[c]{ccc}%
a_{\max} & if & 0\leq t<\tau_{1}\\
-a_{\max} & if & \tau_{1}\leq t\leq\tau_{p}%
\end{array}
\right.  .\label{va}%
\end{equation}

\bigskip It is noteworthy that the value of every parameter, i.e. $\tau_{s}$,
$\tau_{1}$ and $\pm a_{\max}$, are determined by the intended angle of
rotation of the pulse besides Eq.(\ref{e3ea}) and Eq.(\ref{e3eb}). In
addition, the pulse shapes described by Eq.(\ref{vs}) and Eq.(\ref{va}) are
just two of the solutions discussed in
\cite{PhysRevA.77.032315,jpa.41.312005,PhysRevA.78.022315}. In the next section, we will
study which type of pulse, symmetric or asymmetric, is more suitable for the
experimental situation under consideration the small random\ errors\ in\ the rotation\ direction.

\section{Error due to rotation axis ($\eta^{(\epsilon,1)(\tau_{p},0)}$ and
$\eta^{(\epsilon,1)(\tau_{p},1)}$)}

In this section, we consider both $\eta^{(\epsilon,1)(\tau_{p},0)}$
and $\eta^{(\epsilon,1)(\tau_{p},1)}$, which\ are not studied in
\cite{PhysRevA.83.022306}. For short time $\tau_{p}$, one has $\eta^{(\epsilon
,1)(\tau_{p},0)}>\eta^{(\epsilon,1)(\tau_{p},1)}$. Let us first consider the
dominant error $\eta^{(\epsilon,1)(\tau_{p},0)}$ in the error due to rotation
axis. As in section 3, we separate the arbitrary operator of the system
$\Omega^{\prime}$ into two parts%

\begin{equation}
\Omega^{\prime}=\Omega_{a}^{\prime}+\Omega_{c}^{\prime},
\end{equation}
\ with $\Omega_{a}^{\prime}$ and $\Omega_{c}^{\prime}$\ anticommuting and
commuting with $\Omega$, respectively. Hence $\eta^{(\epsilon,1)(\tau_{p},0)}$
in Eq.(\ref{e1}) takes the form of%

\begin{equation}
\eta^{(\epsilon,1)(\tau_{p},0)}=-\Omega\Omega_{a}^{\prime}\eta_{1}%
^{(\epsilon,1)(\tau_{p},0)}-i\Omega_{a}^{\prime}\eta_{2}^{(\epsilon
,1)(\tau_{p},0)}-\Omega\Omega_{c}^{\prime}\eta_{3}^{(\epsilon,1)(\tau_{p}%
,0)}-i\Omega_{c}^{\prime}\eta_{4}^{(\epsilon,1)(\tau_{p},0)},\label{e1e}%
\end{equation}
where%
\begin{equation}
\eta_{1}^{(\epsilon,1)(\tau_{p},0)}=\int\nolimits_{0}^{\tau_{p}}\epsilon
V(t)\sin\left[  \phi_{-}-\psi(t)\right]  dt,\label{e1ea}%
\end{equation}

\begin{equation}
\eta_{2}^{(\epsilon,1)(\tau_{p},0)}=\int\nolimits_{0}^{\tau_{p}}\epsilon
V(t)\cos\left[  \phi_{-}-\psi(t)\right]  dt,\label{e1eb}%
\end{equation}

\begin{equation}
\eta_{3}^{(\epsilon,1)(\tau_{p},0)}=\int\nolimits_{0}^{\tau_{p}}\epsilon
V(t)\sin\left[  \phi_{+}\right]  dt,\label{e1ec}%
\end{equation}

\begin{equation}
\eta_{4}^{(\epsilon,1)(\tau_{p},0)}=\int\nolimits_{0}^{\tau_{p}}\epsilon
V(t)\cos\left[  \phi_{+}\right]  dt,\label{e1ed}%
\end{equation}
with $\phi_{+}$ defined just below Eq.(\ref{e3eb}). It is worth to mention that
the value of $\phi_{+}$ equals half of the intended angle of rotation of the
pulse, such as $\phi_{+}=\pi/2$ for the $\pi$ pulse and $\phi_{+}=\pi/4$ for
the $\pi/2$ pulse.\ Before discussing how pulse shaping reduces the errors
described by Eq.(\ref{e1ea})$\sim$Eq.(\ref{e1ed}), we derive the expressions
of $\phi_{-}-\psi(t)$ for general symmetric and asymmetric pulse shapes

\begin{equation}
\phi_{-}-\psi(t)=\frac{\pi}{2}-2\int_{0}^{t}V(s)ds,
\end{equation}

\bigskip Hence we have%

\begin{eqnarray}
\eta_{1}^{(\epsilon,1)(\tau_{p},0)}  &=& \epsilon\int_{0}^{\tau_{p}}V(t)\cos[2\int_{0}^{t}V(s)ds]dt\nonumber\\
&=&\frac{\epsilon}{2}\int_{0}^{\tau_{p}}\cos[2\int_{0}^{t}V(s)ds]d[2\int_{0}^{t}V(s)ds]\nonumber\\
&=&0,
\end{eqnarray}
\begin{eqnarray}
\eta_{2}^{(\epsilon,1)(\tau_{p},0)}  &=& \epsilon\int_{0}^{\tau_{p}}V(t)\sin[2\int_{0}^{t}V(s)ds]dt\nonumber\\
&=&\frac{\epsilon}{2}\int_{0}^{\tau_{p}}\sin[2\int_{0}^{t}V(s)ds]d[2\int_{0}^{t}V(s)ds]\nonumber\\
&=&\epsilon,
\end{eqnarray}
\begin{equation}
\eta_{3}^{(\epsilon,1)(\tau_{p},0)}=\frac{\pi}{2}\epsilon,
\end{equation}
\begin{equation}
\eta_{4}^{(\epsilon,1)(\tau_{p},0)}=0,
\end{equation}
for both symmetric and asymmetric pulses in terms of the $\pi$ pulse, i.e.
$\Phi_{s}=\Phi_{a}=\pi$. So both pulse shapes have the same effect on
Eq.(\ref{e1ea})\symbol{126}Eq.(\ref{e1ed}). Besides, we have $\frac
{1}{\epsilon}\eta_{2}^{(\epsilon,1)(\tau_{p},0)}\neq0$ and $\eta
_{3}^{(\epsilon,1)(\tau_{p},0)}\neq0$\ no matter what type of composite pulse
ia applied to optimize the coherent control pulse. From this point of view the
impact of errors in the direction of the rotation axis cannot be ignored completely.

\bigskip Now, the analogous procedure is used to obtain%
\begin{eqnarray}
\eta^{(\epsilon,1)(\tau_{p},1)}  =& -i\Omega[(  H_{a}\Omega_{a}^{\prime
}-\Omega_{a}^{\prime}H_{a})+(  H_{c}\Omega_{c}^{\prime
}-\Omega_{c}^{\prime}H_{c})]  \eta_{1}^{(\epsilon,1)(\tau_{p},1)}\nonumber\\
&+[(H_{a}\Omega_{a}^{\prime}-\Omega_{a}^{\prime}H_{a})+(H_{c}\Omega_{c}^{\prime}-\Omega_{c}^{\prime}H_{c})]  \eta_{2}%
^{(\epsilon,1)(\tau_{p},1)}\nonumber\\
& -i\Omega2\left(  H_{a}\Omega_{c}^{\prime}+H_{c}\Omega_{a}^{\prime}\right)
\eta_{3}^{(\epsilon,1)(\tau_{p},1)}+2\left(  H_{a}\Omega_{c}^{\prime}%
+H_{c}\Omega_{a}^{\prime}\right)  \eta_{4}^{(\epsilon,1)(\tau_{p}%
,1)},\label{e2e}%
\end{eqnarray}
where%
\begin{equation}
\eta_{1}^{(\epsilon,1)(\tau_{p},1)}=\epsilon\int\nolimits_{0}^{\tau_{p}%
}(t-\tau_{s})V(t)\sin\left[  \phi_{+}\right]  dt,\label{e2ea}%
\end{equation}

\begin{equation}
\eta_{2}^{(\epsilon,1)(\tau_{p},1)}=\epsilon\int\nolimits_{0}^{\tau_{p}%
}(t-\tau_{s})V(t)\cos\left[  \phi_{+}\right]  dt,\label{e2eb}%
\end{equation}

\begin{equation}
\eta_{3}^{(\epsilon,1)(\tau_{p},1)}=\epsilon\int\nolimits_{0}^{\tau_{p}%
}(t-\tau_{s})V(t)\sin\left[  \phi_{-}-\psi(t)\right]  dt,\label{e2ec}%
\end{equation}

\begin{equation}
\eta_{4}^{(\epsilon,1)(\tau_{p},1)}=\epsilon\int\nolimits_{0}^{\tau_{p}%
}(t-\tau_{s})V(t)\cos\left[  \phi_{-}-\psi(t)\right]  dt,\label{e2ed}%
\end{equation}
Hence, we find that both symmetric and asymmetric pulses (see
Eq.(\ref{vs}) and Eq.(\ref{va})) can make terms $\eta_{3}^{(\epsilon
,1)(\tau_{p},1)}$ and $\eta_{4}^{(\epsilon,1)(\tau_{p},1)}$\ vanish. In
addition, we obtain $\eta_{2}^{(\epsilon,1)(\tau_{p},1)}=0$ for the $\pi$
pulse, i.e. $\phi_{+}=\pi/2$. Next, we look at Eq.(\ref{e2ea}). For symmetric
pulse shape, Eq.(\ref{e2ea}) gives%

\begin{eqnarray}
\frac{1}{\epsilon}\eta_{1,s}^{(\epsilon,1)(\tau_{p},1)} & =\int^{\tau_{s}}_{0}(t-\tau_{s})V(t)dt+\int^{2\tau_{s}}_{\tau_{s}}(t-\tau_{s})V(t)dt\nonumber\\
& =0,\label{s67}%
\end{eqnarray}
we have use the relation $V(t)=V(2\tau_{s}-t)$ in the last line of
Eq.(\ref{s67}). It should be noted that we have $\eta_{1,s}^{(\epsilon,1)(\tau_{p},1)}=0$ for all symmetric $\pi$ pulse, including the specific shaped pulse given in Eq.(\ref{vs}). For asymmetric pulse shape described by Eq.(\ref{va}), this error term is in the form of
\bigskip%
\begin{equation}
\frac{1}{\epsilon}\eta_{1,a}^{(\epsilon,1)(\tau_{p},1)}=a_{\max}(\tau_{1}^{2}-\frac{\tau_{p}%
^{2}}{2}-2\tau_{1}\tau_{s}+\tau_{p}\tau_{s}).\label{a67}%
\end{equation}
In spite of our intensive search we have not succeeded in proving $\tau
_{1}^{2}-\tau_{p}^{2}/2-2\tau_{1}\tau_{s}+\tau_{p}\tau_{s}\neq0$. However, we
argue that at least some asymmetric pulse shapes cannot make $\eta
_{1,a}^{(\epsilon,1)(\tau_{p},1)}$ vanish. Take $\tau_{1}=(2n+1)\tau_{p}%
/(4n)$, $\tau_{s}=\tau_{p}[1/2+(-1)^{n}/(2n\pi)]$ and $a_{\max}=\pi n/\tau
_{p}$, where $n$ is a positive integer, for instance, we get%
\begin{equation}
\frac{1}{\epsilon}\eta_{1,a}^{(\epsilon,1)(\tau_{p},1)}=\frac{[-4(-1)^{n}+\pi-4n^{2}\pi]\tau
_{p}}{16n}\neq0.
\end{equation}

Therefore, the symmetric pulse shape eliminates the errors more effectively
than the asymmetric one does in terms of Eq.(\ref{e2ea}) for the specific
choice of pulse shapes.

Finally, the results discussed in section 3 and 4 are listed in Table~\ref{1}. As shown in Table~\ref{1}, we should note that except the term $\eta_{1}^{(\epsilon,1)(\tau_{p},1)}$ for the specific choice of asymmetric pulse, the rest results are not dependent on the specific form of pulse shape (given by Eq.(\ref{vs}) and Eq.(\ref{va})), that is, we get the rest results from all symmetric or asymmetric pulse.

\begin{table}
\caption{\label{1}The comparison between symmetric pulse (SP) and asymmetric pulse (AP) for all terms discussed in section 3. According to the additional condition (AC) and results, the terms have been divided into 4 groups. With the additional condition, we will have the result, $``=0"$ or $``\neq0"$, for the terms. The result $``=0"$ or $``\neq0"$ means that the corresponding term is zero or nonzero. The additional condition ``no AC" means there is no additional condition. $``\Phi=\pi"$ implies that the pulse considered here is a $\pi$ pulse. And $``\triangle"$ indicates that we have the corresponding results for all symmetric pulse and just for the specific choice of asymmetric pulse. }
\begin{indented}
\lineup
\item[]\begin{tabular}{@{}*{4}{l}}
\br
$\0\0\0\0\0\0\0\0Terms$&$SP$&$AP$&\m$AC$\cr
\mr
$\eta_{1}^{(\tau_{p},1)}$,$\eta_{2}^{(\tau_{p},1)}$,$\eta_{3}^{(\epsilon,1)(\tau_{p},1)}$,$\eta_{4}^{(\epsilon,1)(\tau_{p},1)}$&$=0$  &$=0$&no AC\cr
$\eta_{1}^{(\epsilon,1)(\tau_{p},0)}$,$\eta_{4}^{(\epsilon,1)(\tau_{p},0)}$,$\eta_{2}^{(\epsilon,1)(\tau_{p},1)}$&$=0$&$=0$&$\Phi=\pi$\cr
$\eta_{2}^{(\epsilon,1)(\tau_{p},0)}$,$\eta_{3}^{(\epsilon,1)(\tau_{p},0)}$&$\neq0$&$\neq0$&$\Phi=\pi$\cr
$\eta_{1}^{(\epsilon,1)(\tau_{p},1)}$&$=0$&$\neq0$&$\triangle$\cr
\br
\end{tabular}
\end{indented}
\end{table}

\section{Conclusions}
In this paper, we follow the method outlined in \cite{PhysRevA.83.022306} to
study a finite-amplitude short coherent control pulse with small random errors
in its direction. Note that the errors in direction is neglected in
\cite{PhysRevA.83.022306}. The error $\eta^{(\tau_{p},1)}$, which comes from pulse
duration $\tau_{p}$ only, is reviewed first and the optimized pulse shapes for
a single-qubit flip control can be used to eliminate this error as
expected\cite{PhysRevA.77.032315,jpa.41.312005,PhysRevA.78.022315}.\ Secondly, we consider
the error terms $\eta^{(\epsilon,1)(\tau_{p},0)}$ and $\eta^{(\epsilon
,1)(\tau_{p},1)}$, which originate from deviation in rotation axis. The result
shows that the term $\eta^{(\epsilon,1)(\tau_{p},0)}$ can not be removed
completely by either symmetric or asymmetric shaped pulse and both types of
pulse have the same effect in view of this term. As for the term
$\eta^{(\epsilon,1)(\tau_{p},1)}$, we find that any symmetric pulse shaping
can make it vanish while the specific choice of asymmetric one can not.\ It should be mentioned
that symmetric pulse shape performs better than asymmetric pulse just for the
specific pulse shapes which we choose. This work can be useful for
experimental implementations of kinds of DD protocols, and for deeper
understanding of a short coherent control pulse which is subject\ to random
errors in its directions.

Further work should verify the results in specific models in order to
understand better how important the pulse shape in approximating $\delta
$-shaped pulse really are. In addition, the more generic situation with pulse
errors, e.g. pulse length errors, amplitude errors or an off-resonance error,
is much more complex and further investigation is needed.

\section*{Acknowledgements}

This work is supported by the National Natural Science Foundation of China under Grant No.60977042.

\section*{References}\label{refs}
\bibliographystyle{unsrt}
\bibliography{jpb}

\end{document}